\documentclass[conference,a4paper]{IEEEtran}

\usepackage{cite}
\usepackage{amsmath,amssymb}
\usepackage{graphicx}
\usepackage{booktabs}
\usepackage{url}

\begin{document}

\title{saanoTTS: The Smallest Real-Time Neural TTS on a
General-Purpose Microcontroller}

\author{\IEEEauthorblockN{Ashish Thapa}
\IEEEauthorblockA{Ampixa Labs \\
\url{ashish@ampixa.com}}}

\maketitle

\begin{abstract}
This paper describes an audited neural text-to-speech stack that runs from
phoneme IDs to 22.05-kHz PCM on general-purpose microcontrollers.  Its
deployed graph has 567{,}008 parameters, and its two int8 blobs occupy
679{,}832 bytes.  On an ESP32-S3, the complete
duration--acoustic--inverse-STFT path generates 4.54 s of speech in 1.02 s
($0.22\times$ real time) without a neural accelerator.  The same portable C
core runs offline at $5.72\times$ real time on an FPU-less ESP32-C3.  To our
knowledge, this is the smallest complete phoneme-to-waveform neural TTS graph
demonstrated in real time on a general-purpose microcontroller without a
neural accelerator.  We
derive the students from the conditional-VAE objective of their Piper/VITS
teachers and state the duration, latent-interface, waveform, adversarial, and
joint-distillation losses used in training.  The size and speed come with an
audible cost: on unseen text, the embedded stack distilled from
\texttt{en\_US-kristin-medium} scores 2.54 SCOREQ and 2.80 UTMOS, compared
with 4.68 and 4.42 for its teacher.  A separate English quality package uses
the stronger \texttt{en\_US-amy-medium} teacher.  Its 1{,}454{,}284-parameter
Pareto point scores 4.13 SCOREQ and 4.10 UTMOS; a 1{,}834{,}380-parameter
variant scores 4.16 SCOREQ.  A controlled capacity study with Kristin
identifies the decoder, rather than the output representation, as the main
constraint.  Two evaluation failures also affected the work: a narrow,
templated test set overstated one early student's SCOREQ by 1.35, and
aggregate quality predictors missed a sibilant failure that was evident in
listening and in a phoneme-resolved spectral probe.  Checksums cover the
reported model blobs, runtime ports, and golden vectors.
\end{abstract}


\section{Introduction}
\label{sec:intro}

Most neural TTS systems pair an acoustic model with a neural
vocoder~\cite{ren2021fastspeech2,kong2020hifigan}.  Piper~\cite{piper}, built
on VITS~\cite{kim2021vits}, provides strong open teacher voices.  Its roughly
15-M-parameter models, however, are intended for Linux-class devices.  For a
microcontroller target, parameter count answers only part of the question.
The model must also retain useful quality on unseen text and finish synthesis
within the device's compute and memory limits.

Knowledge distillation offers one route to that target because it transfers a
teacher's behavior instead of only pruning its tensors~\cite{hinton2015distilling}.
Published embedded speech results do not all use the same system boundary.
The TinyTTS hardware module, for example, uses an Ethos-U55 neural
accelerator, whereas TinyVocos measures a mel-to-waveform vocoder rather than
a complete text-to-speech stack~\cite{tinytts_hw,tinyvocos}.  SlimTTS reports
an even smaller 562k-parameter phoneme-to-mel plus HiFi-GAN pipeline, but
positions it for future custom hardware rather than reporting execution on a
physical microcontroller~\cite{debeer2026slimtts}.  Here we time the full
neural path, without an NPU, on physical microcontrollers.

The central deployment result is the 40-dimensional \emph{c-line}.  It joins
a duration student, an acoustic student that predicts the decoder contract
directly, and a frame-domain inverse-STFT decoder, for a total of exactly
567{,}008 inference parameters.  A later 600{,}097-parameter \emph{z-line}
improved the automatic quality score, but its waveform-domain decoder requires
522 MMAC/s.  We therefore treat it as an ablation rather than as an MCU
deployment.  Both models, along with a 1.396-M-parameter diagnostic package,
were distilled from Kristin and can be compared directly.  The current
quality release was distilled from Amy: its 1.454-M model is the default
English voice, and a 1.834-M variant has the highest student SCOREQ.  Results
from the two teachers are kept separate throughout the paper.

Our contributions are:
\begin{enumerate}
\item a reproducible derivation from the VITS conditional VAE to three
deterministic students joined by an explicit latent interface;
\item an evaluation protocol that separates teacher, decoder, acoustic,
full-stack, and phoneme-class effects and exposes test-set inflation;
\item a controlled capacity study using Kristin, followed by a separate Amy
replication; and
\item an NPU-free deployment measured at $0.22\times$ real time on ESP32-S3
and $5.72\times$ on ESP32-C3, with both outputs checked against floating-point
golden audio at correlation of at least 0.98.
\end{enumerate}

The code, model manifests, exact binary fixtures, and evaluation tools are in
the saanoTTS repository~\cite{saanottsrepo}.

\section{From Each Teacher VAE to Deterministic Students}
\label{sec:math}

Let $x$ denote a phoneme sequence, $A$ its monotonic alignment, $y$ the
waveform, and $z$ the latent presented to the VITS generator.  Each voice
$v$ has its own teacher, frontend map, and student weights; for readability,
we omit the superscript $v$.  VITS is a conditional variational autoencoder
with posterior $q_\phi(z\mid y)$ and text-conditioned prior
$p_\theta(z\mid x,A)$.  Ignoring the separate duration and adversarial terms,
the evidence lower bound maximized by the teacher is
\begin{align}
\mathcal{L}_{\mathrm{VAE}}={}&
 \mathbb{E}_{q_\phi(z\mid y)}[\log p_\theta(y\mid z)] \nonumber\\
 &-D_{\mathrm{KL}}\!\left(q_\phi(z\mid y)\,\|\,
 p_\theta(z\mid x,A)\right).
\label{eq:vae}
\end{align}
The normalizing flow gives the VITS prior its expressiveness, and the
generator maps a sample back to waveform space~\cite{kim2021vits}.  Our
students do not form a second VAE.  We freeze the teacher and run deterministic
inference, retaining its durations $d^T$, one fixed prior sample $z^T$, and
waveform $y^T$.  Although the posterior $q_\phi$ is part of VAE training, the
distillation labels come from the text-conditioned prior.  For every promoted
pack, label generation sets the noise and duration-noise scales to zero and
the length scale to one.  Each teacher therefore supplies a consistent set of
labels, and no pack mixes labels from different voices.

\begin{figure*}[t]
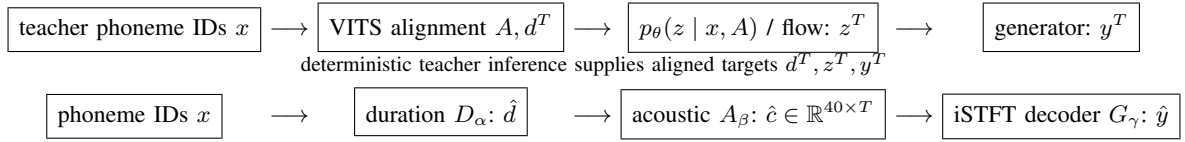

\centering
\small
\setlength{\fboxsep}{4pt}
\begin{tabular}{c@{\ $\longrightarrow$\ }c@{\ $\longrightarrow$\ }c@{\ $\longrightarrow$\ }c}
\fbox{teacher phoneme IDs $x$} &
\fbox{VITS alignment $A,d^T$} &
\fbox{$p_\theta(z\mid x,A)$ / flow: $z^T$} &
\fbox{generator: $y^T$} \\
\multicolumn{4}{c}{\footnotesize deterministic teacher inference supplies aligned targets $d^T,z^T,y^T$}\\[3pt]
\fbox{phoneme IDs $x$} &
\fbox{duration $D_\alpha$: $\hat d$} &
\fbox{acoustic $A_\beta$: $\hat c\in\mathbb{R}^{40\times T}$} &
\fbox{iSTFT decoder $G_\gamma$: $\hat y$}
\end{tabular}
\caption{Teacher-coherent supervision and the packaged embedded inference
path.  The teacher posterior in~(\ref{eq:vae}) is used during VAE training;
distillation labels follow the text prior shown here.  During decoder training
a learned encoder produces
$c^T=E_\rho(z^T)$; at deployment the acoustic student predicts $c$ directly,
so $E_\rho$ is not counted or executed.}
\label{fig:pipeline}
\end{figure*}

\subsection{Duration and latent-interface distillation}

The duration network emits log-duration $\hat\ell_i$.  Deployment applies a
per-voice calibrated length scale $s_v$ and clips the result:
\begin{equation*}
r_i=\max(1,e^{\hat\ell_i}),\qquad
\hat d_i=\operatorname{clip}_{[1,80]}\!\left(\operatorname{round}(s_vr_i)\right).
\end{equation*}
Here $s_v=1.08$ for the released English quality voices and 1.16 for the
Vietnamese and Indonesian releases.
The objective combines token accuracy with utterance-length conservation:
\begin{equation}
\mathcal{L}_{d}=\operatorname{Huber}_{0.25}
(\hat\ell,\log d^T)+\lambda_T
\left[\log\!\sum_i r_i-\log\!\sum_i d_i^T\right]^2.
\label{eq:dur}
\end{equation}

For the embedded c-line, a training-only encoder contracts the 192-channel
teacher latent to $c^T=E_\rho(z^T)\in\mathbb{R}^{40\times T}$.  The acoustic
student predicts $\hat c=A_\beta(x,\hat d)$ directly.  Let
$\mathcal N_T(u)=(u-\mu_T)/\sigma_T$ denote channel normalization by the
teacher-pack statistics.  Its loss is the weighted sum implemented by the
trainer:
\begin{align}
\mathcal{L}_{c}={}&\|\hat c-c^T\|_1
+\lambda_2\|\hat c-c^T\|_2^2
+\lambda_n\|\mathcal N_T(\hat c)-\mathcal N_T(c^T)\|_1 \nonumber\\
&+\lambda_\Delta\|\Delta\hat c-\Delta c^T\|_1+\lambda_s\mathcal L_{\rm stat},
\nonumber\\
\mathcal L_{\rm stat}={}&
\left\|\frac{\mu(\hat c)-\mu(c^T)}{\sigma_T}\right\|_1
+\left\|\frac{\sigma(\hat c)-\sigma(c^T)}{\sigma_T}\right\|_1.
\label{eq:latent}
\end{align}
The channel-normalized terms keep high-variance latent channels from
dominating the objective.  In the Kristin diagnostic package and the Amy
quality packages, the full 192-channel $z$ replaces $c$.  These packages also
use a hinge adversary to discourage conditional mean collapse:
\begin{align}
\mathcal L_D^z={}&\mathbb E[(1-D_z(z^T))_+]
+\mathbb E[(1+D_z(\hat z))_+], \nonumber\\
\mathcal L_{z,\mathrm{adv}}={}&-\mathbb E[D_z(\hat z)].
\end{align}

\subsection{Decoder and joint distillation}

During training, the decoder receives both exact teacher contracts and
contracts predicted by the student.  Its multi-resolution STFT term uses FFT
sizes 512, 1024, and 2048, with hops 128, 256, and 512.  Let $\mathcal R$
denote these resolution pairs and $\hat y=G_\gamma(\tilde c)$.  The core
generator objective is
\begin{align}
\mathcal{L}_{G}(\hat y,y^T)={}&\lambda_w\|\hat y-y^T\|_1
+\frac{\lambda_{S}}{|\mathcal R|}\sum_{(n,h)\in\mathcal R}
\ell_{n,h} \nonumber\\
&+\lambda_A\mathcal{L}_{\mathrm{adv}}+\lambda_F\mathcal{L}_{\mathrm{FM}},
\nonumber\\
\ell_{n,h}={}&\|\log(1+|S_{n,h}(\hat y)|)
-\log(1+|S_{n,h}(y^T)|)\|_1,
\label{eq:decoder}
\end{align}
where $\tilde c=c^T$ or $\hat c$ under predicted-code mixing,
$\mathcal{L}_{\mathrm{adv}}=\mathcal{L}_{\mathrm{LSGAN}}(D(\Delta\hat y))$ is
the least-squares generator loss~\cite{mao2017lsgan}, and
$\mathcal{L}_{\mathrm{FM}}$ matches discriminator features of $\Delta\hat y$
and $\Delta y^T$.  Multi-resolution spectral loss follows the same motivation as
Parallel WaveGAN~\cite{yamamoto2020parallelwavegan}.  The final c-line stage
uses a first-difference discriminator and updates acoustic and decoder together:
\begin{align}
\hat y={}&G_\gamma(A_\beta(x,\hat d)), \nonumber\\
\mathcal{L}_{\mathrm{joint}}={}&\mathcal{L}_{G}(\hat y,y^T)
+\lambda_c\|A_\beta(x,\hat d)-c^T\|_1.
\label{eq:joint}
\end{align}
The packaged run uses
\begin{equation*}
(\lambda_w,\lambda_S,\lambda_A,\lambda_F,\lambda_c)
=(0.1,0.5,0.025,0.25,0.5).
\end{equation*}
The final term anchors the interface while allowing the decoder to adapt to
errors from the acoustic student.

\section{Artifacts and Evaluation Protocol}
\label{sec:protocol}

The project uses a separate teacher for each voice.  The controlled MCU study
uses \texttt{en\_US-kristin-medium}, whereas the current English quality
release uses \texttt{en\_US-amy-medium}.  The Vietnamese and Indonesian
packages use \texttt{vi\_VN-vais1000-medium} and
\texttt{id\_ID-news\_tts-medium}, respectively.  Each package is trained from
one teacher tuple and ships its own weights and matching Piper/eSpeak
phoneme-ID map; only the training recipe is shared across
languages~\cite{espeakng}.  All four voices generate 22.05-kHz audio.

Our primary English test set, \emph{diverse24}, consists of 24 LJSpeech
validation sentences that do not occur in training~\cite{ito2017ljspeech}.
Metrics are computed on raw full-stack output.  They include the synthetic,
no-reference SCOREQ measure~\cite{ragano2024scoreq},
UTMOS~\cite{saeki2022utmos}, DNSMOS signal quality~\cite{reddy2021dnsmos}, and
normalized Whisper-small WER~\cite{radford2023whisper}.  We also listened to
the output of every promoted model.

Table~\ref{tab:artifacts} distinguishes three systems that were conflated in
earlier drafts.  Counts include live neural tensors but exclude the external
text frontend and the training-only discriminators.  The embedded count also
excludes the 14{,}952-parameter $z\!\to\!c$ encoder because it is used only in
training; at deployment, the acoustic model already emits $c$.  Conversely,
the deployed acoustic vocabulary has 157 entries, 30 more than the research
configuration, and adds 1{,}440 embedding parameters.  The shipped graph
therefore has 567{,}008 parameters rather than the earlier count of 565{,}568.
Its two blobs occupy 280{,}288 and 399{,}544 bytes.  Padding, floating-point
biases, and per-channel scales account for the difference between parameter
count and binary size.

\begin{table}[t]
\caption{Audited operating points from the same Kristin teacher.  The first
and third rows are current artifacts; R8 is a historical ablation.}
\label{tab:artifacts}
\centering
\small
\begin{tabular}{@{}lrrr@{}}
\toprule
System & Duration & Acoustic & Decoder / total \\
\midrule
embedded c-line & 36{,}164 & 199{,}536 & 331{,}308 / \textbf{567{,}008} \\
R8 z-line ablation & 36{,}164 & 206{,}408 & 357{,}525 / 600{,}097 \\
quality z-line & 36{,}164 & 359{,}112 & 1{,}000{,}875 / \textbf{1{,}396{,}151} \\
\bottomrule
\end{tabular}
\end{table}

Table~\ref{tab:voices} records the teacher associated with each current
artifact, as listed in the package manifests.  On the common English gate,
Amy is the strongest Piper teacher we evaluated (SCOREQ 4.71).  Its 1.834-M
student has the highest student score, 4.16, while the smaller 1.454-M package
is the default release.  For Vietnamese and Indonesian, we report only the
ratio to the corresponding teacher because absolute SCOREQ values are not
calibrated for comparisons across languages.

\begin{table}[t]
\caption{Audited teacher-to-package mapping.  Parameters are manifest totals;
quality evidence is not compared across languages.}
\label{tab:voices}
\centering
\footnotesize
\setlength{\tabcolsep}{4.5pt}
\begin{tabular}{@{}lllrl@{}}
\toprule
Language & Teacher voice & Package & Params & Evidence \\
\midrule
English & Kristin & MCU fixture & 567{,}008 & S3 measured \\
English & Amy & Pareto & 1{,}454{,}284 & SCOREQ 4.13 \\
English & Amy & champion & 1{,}834{,}380 & SCOREQ 4.16 \\
Vietnamese & Vais1000 & quality & 1{,}565{,}484 & 0.82$\times$ teacher \\
Indonesian & NewsTTS & quality & 1{,}562{,}124 & 0.82$\times$ teacher \\
\bottomrule
\end{tabular}
\end{table}

The generated deployment headers fix the topology as well as the weights.
The duration model has width 32 and three kernel-5 residual blocks.  The
acoustic model has width 48, three token blocks, five frame blocks, and 40
output channels.  The decoder maps this contract to width 76 and applies five
kernel-7 blocks with 304-channel pointwise expansions and rank-12
conditioning.  It then predicts 513 magnitude bins and 1{,}026 phase
coordinates for a 1024-point iSTFT with hop 256.  Training the promoted c-line
acoustic model used 14{,}343 teacher-labelled text rows; its run reports
record the exact seeds and flags.

\begin{figure}[t]
\centering
\setlength{\fboxsep}{5pt}
\begin{minipage}{0.46\columnwidth}
\centering
\fbox{\begin{minipage}{0.88\linewidth}\centering
\textbf{Embedded tier}\\
Kristin teacher\\
567{,}008 parameters\\
680 KB int8\\
SCOREQ 2.54\\
ESP32-S3: 0.22 RT
\end{minipage}}
\end{minipage}\hfill
\begin{minipage}{0.46\columnwidth}
\centering
\fbox{\begin{minipage}{0.88\linewidth}\centering
\textbf{Quality tier}\\
Amy teacher\\
1{,}454{,}284 parameters\\
2.91 MB fp16\\
SCOREQ 4.13\\
browser / desktop
\end{minipage}}
\end{minipage}
\caption{The current deployment choices.  Because the tiers use different
teachers, this is a product frontier, not a controlled architecture ablation;
the Kristin-only comparison in Table~\ref{tab:artifacts} supplies that control.}
\label{fig:tiers}
\end{figure}

\section{What the Evaluation Revealed}
\label{sec:findings}

\subsection{How a narrow test set misled us}

Our first held-out set contained 12 templated sentences.  On that set, an
early 565{,}568-parameter c-line scored 3.07, which initially appeared
promising.  The result did not survive a change of text distribution: the
same model scored only 1.72 on diverse24.  The teacher changed much less,
from 4.82 to 4.68.  In other words, the templated set overstated the student's
SCOREQ by 1.35.  Expanding the teacher-labelled training set from 512 to
14{,}343 rows raised the diverse24 result to 2.54.  We use that value for the
current embedded artifact, rather than its narrow-set score of 3.04.

We tested the interface separately with 512 training rows and matched model
budgets.  A merged text-to-waveform student memorized the training rows but
failed on held-out text, scoring 1.06 SCOREQ with 0.28 WER.  A factored model
with the explicit 40-dimensional interface scored 2.95 with zero WER on the
same 12 rows.  At this data scale, the compact merged network did not recover
the generalization already present in the teacher latent.

\begin{table}[t]
\caption{English quality on unseen diverse24.  UTMOS and DNS-SIG were measured
on the same 24 renders.  Kristin rows form the controlled MCU study; Amy rows
are a separate teacher replication.}
\label{tab:quality}
\centering
\footnotesize
\setlength{\tabcolsep}{4.5pt}
\begin{tabular}{@{}lrrrr@{}}
\toprule
System & Params & SCOREQ & UTMOS & DNS-SIG \\
\midrule
embedded fixture & 0.567 M & 2.54 & 2.80 & 3.26 \\
R8 ablation & 0.600 M & 2.94 & --- & --- \\
Kristin diagnostic ($\beta=0$) & 1.396 M & 4.09 & 3.98 & 3.58 \\
Kristin diagnostic ($\beta=6$) & 1.396 M & 3.92 & 3.89 & 3.57 \\
Amy Pareto package & 1.454 M & 4.13 & \textbf{4.10} & 3.61 \\
Amy champion package & 1.834 M & \textbf{4.16} & 4.06 & 3.61 \\
open TinyTTS & 1.220 M$^*$ & 3.94 & 3.65 & \textbf{3.62} \\
Kristin teacher & $\sim$15.7 M & 4.68 & 4.42 & 3.59 \\
Amy teacher & $\sim$15 M & 4.71 & 4.47 & 3.65 \\
\bottomrule
\multicolumn{5}{@{}l@{}}{$^*$Folded ONNX initializers; PyTorch state is 1.62 M.}
\end{tabular}
\end{table}

For open TinyTTS, Table~\ref{tab:quality} uses the conservative inference count
of 1{,}219{,}596 parameters~\cite{tinytts}.  By this count, neither of our
English quality packages is smaller.  The Amy Pareto package scores higher
than TinyTTS on SCOREQ and UTMOS, but slightly lower on DNSMOS-SIG (3.61 versus
3.62).  This disagreement is one reason we do not treat any single predictor
as ground truth.  Both Piper teachers, as well as the much larger Kokoro
class, score above our students; we do not claim state-of-the-art
quality~\cite{kokoro}.  With the committed duration path, normalized WER is
14.8\% for the embedded fixture and 11.7\% for R8.  We did not record WER for
every subsequent comparison.

\subsection{Decoder capacity defines the quality tier}

To locate the remaining error, we passed teacher and student latents through
teacher and student decoders.  On diverse24, the teacher latent $z$ scores
4.68 through the teacher decoder.  Replacing it with the jointly trained
acoustic model's latent lowers the score to 3.70.  In the other direction,
teacher $z$ scores 3.20 through the 357k R8 decoder and 4.13 through the
1.0-M teacher-initialized decoder.  Thus the decoder gap at 357k is 1.49,
larger than the acoustic gap of 0.98.  Joint co-adaptation recovers some of the
combined loss, but not enough to remove this capacity boundary.

The deployment choice follows from the same comparison.  R8 uses a 357k
waveform-domain transposed-convolution decoder and requires 522 MMAC/s.  The
331k embedded decoder evaluates its learned convolutions at 86 frames/s and
uses a 1024-point inverse STFT for sample-rate synthesis, reducing the cost to
28 MMAC/s.  This representation round-trips teacher spectra within
$\pm0.005$ SCOREQ.  Its lower student score therefore comes from prediction
and model capacity, not from an inverse-STFT ceiling.

\subsection{Aggregate scores miss phoneme-class failure}

The 1.396-M Kristin diagnostic package exposed a failure that the aggregate
scores did not: despite SCOREQ 4.09 and UTMOS 3.98, its sibilants sounded
whistly.  We used the duration alignment to measure 2--8-kHz spectral flatness
by phoneme class for the teacher, an oracle-decoder path (teacher latent
through our decoder), and the full student.  For sibilants, flatness fell from
0.689 in the teacher to 0.590 in the student, while the oracle-decoder path
reached 0.693.  The acoustic student, rather than the decoder, had smoothed out
the stochastic frication.

At inference we restore only that missing variance:
\begin{equation}
\tilde z_{t,k}=\hat z_{t,k}+\mathbf{1}[x_t\in\mathcal S]\,
\beta\,\sigma^T_k\epsilon_{t,k},\qquad \epsilon_{t,k}\sim\mathcal N(0,1),
\label{eq:sibilant}
\end{equation}
where $\mathcal S=\{/s,\,\mathrm{sh},\,z,\,\mathrm{zh}/\}$ and $\sigma^T_k$
is the teacher's calibrated standard deviation for latent channel $k$.
We selected $\beta=6$ by listening.  It raises sibilant flatness to 0.655 but
reduces aggregate SCOREQ to 3.92 (Table~\ref{tab:quality}).  Both values are
reported because the repair improves the audible defect even though the
aggregate predictor moves in the opposite direction.

\section{Measured Microcontroller Deployment}
\label{sec:deployment}

We export the embedded graph with symmetric int8 weights quantized per output
channel, while activations are quantized per frame.  Embeddings, normalization
affines, and the inverse-STFT support code remain in floating point.  The
portable C99 core uses a caller-owned arena.  A small port interface provides
the int8/int16 dot and matrix-vector kernels, memory-residency decisions,
parallel execution, scratch selection, and timing.  To pass the golden test,
each port must reproduce the embedded reference audio with Pearson correlation
of at least 0.98.

Measurements were taken on a dual-core 240-MHz Xtensa LX7 ESP32-S3 and a
single-core 160-MHz RV32IMC ESP32-C3.  Wall time includes duration prediction,
acoustic inference, decoding, and PCM generation, but not playback.  The S3
uses the 100{,}096-sample golden utterance (4.539 s).  The C3's smaller arena
requires a 34{,}304-sample utterance (1.556 s).  Real-time factor (RTF)
normalizes this difference in length; it measures throughput, not latency to
the first audio sample.

The complete graph performs about 45 MMAC for each second of output audio.  On
the ESP32-S3, SIMD operands have to be staged in internal SRAM; reading vectors
through the flash mapping produced plausible output but only 0.011 correlation
with the reference.  Hand-written PIE matrix-vector kernels, a half-size real
iFFT, frozen normalization, fast math, dual-core column splits, and a periodic
overlap-add envelope reduced compute time for the same 4.54-s utterance from
6.685 to 1.021 s (Table~\ref{tab:runtime}).  The final output correlates 0.985
with the floating-point reference, and peak arena use is about 289 KB.

\begin{table}[t]
\caption{Measured execution of the packaged 567{,}008-parameter graph.  RTF is
compute time divided by audio duration; lower is faster.}
\label{tab:runtime}
\centering
\footnotesize
\begin{tabular}{@{}lrrr@{}}
\toprule
Target / build & audio (s) & compute (s) & RTF \\
\midrule
S3 first end-to-end & 4.54 & 6.685 & 1.47 \\
S3 SIMD + residency & 4.54 & 3.161 & 0.70 \\
S3 PIE + iFFT + dual core & 4.54 & 1.405 & 0.31 \\
\textbf{S3 final} & 4.54 & \textbf{1.021} & \textbf{0.22} \\
ESP32-C3 first boot & 1.56 & 18.680 & 12.01 \\
\textbf{C3 integer pipeline} & 1.56 & \textbf{8.900} & \textbf{5.72} \\
\bottomrule
\end{tabular}
\end{table}

\begin{figure}[t]
\centering
\setlength{\unitlength}{1mm}
\begin{picture}(86,34)
  \put(0,29){\makebox(24,0)[r]{\scriptsize Host C (M3 Pro)}}
  \put(0,24){\makebox(24,0)[r]{\scriptsize WASM / Node}}
  \put(0,19){\makebox(24,0)[r]{\scriptsize ESP32-S3 core}}
  \put(0,14){\makebox(24,0)[r]{\scriptsize ESP32-C3 core}}
  \put(27,8){\line(1,0){55}}
  \multiput(68.25,8)(0,1.2){20}{\line(0,1){0.55}}
  \put(68.25,32){\makebox(0,0){\scriptsize RTF$=1$}}
  \put(27,7){\line(0,1){2}}
  \put(40.75,7){\line(0,1){2}}
  \put(54.5,7){\line(0,1){2}}
  \put(68.25,7){\line(0,1){2}}
  \put(82,7){\line(0,1){2}}
  \put(27,4.8){\makebox(0,0){\scriptsize .001}}
  \put(40.75,4.8){\makebox(0,0){\scriptsize .01}}
  \put(54.5,4.8){\makebox(0,0){\scriptsize .1}}
  \put(68.25,4.8){\makebox(0,0){\scriptsize 1}}
  \put(82,4.8){\makebox(0,0){\scriptsize 10}}
  \put(54.5,1.8){\makebox(0,0){\scriptsize real-time factor (log scale)}}
  \put(35.5,29){\circle{1.8}}
  \put(37.2,29){\makebox(0,0)[l]{\scriptsize 0.0042}}
  \put(44.1,24){\circle{1.8}}
  \put(45.8,24){\makebox(0,0)[l]{\scriptsize 0.017}}
  \put(59.3,19){\circle*{1.8}}
  \put(61.0,19){\makebox(0,0)[l]{\scriptsize 0.22}}
  \put(78.7,14){\circle*{1.8}}
  \put(77.0,14){\makebox(0,0)[r]{\scriptsize 5.72}}
\end{picture}
\caption{Final synthesis throughput across checked execution targets.  Raw
compute/audio times are 6.50 ms/1.556 s for host C and 0.079/4.539 s for
WASM/Node (five-run medians on the M3 Pro), 1.021/4.539 s for S3, and
8.900/1.556 s for C3.  Filled markers are physical boards; open markers are
desktop portability checks, not MCU results.}
\label{fig:targets}
\end{figure}
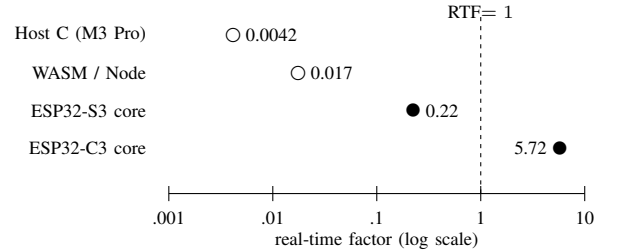

The bottleneck is different on the scalar, FPU-less RV32IMC C3: floating-point
support code, rather than int8 MACs, dominates runtime.  Replacing it with a
block-scaled int32 iFFT, activation tables, integer requantization and phase,
and fixed-point magnitude reduces RTF from 12.01 to 5.72 while retaining the
0.98 correlation threshold.  The C3 nevertheless remains an offline target.
The same weights and core pass the WebAssembly golden test at correlation
0.987.  Host and WebAssembly timings appear in Fig.~\ref{fig:targets} only to
show portability.  RVV kernels pass 200/200 bit-exact tests under QEMU, and
the RP2040 port compiles, but neither port has a synthesis-time measurement
from physical silicon.

\section{Limitations}
\label{sec:limits}

Several limitations matter when interpreting these results.  First, the
embedded quality gap remains large and audible: its SCOREQ is 2.54, compared
with 4.68 for the Kristin teacher.  The 1.396-M Kristin diagnostic model
closes much of this gap, but its sibilant repair has not been ported to the
embedded c-line.  The Amy quality packages use another teacher and were not
run on the MCU, so they are not evidence for controlled MCU scaling.  All
reported hardware quality and speed measurements therefore remain
Kristin-specific.

Second, the Vietnamese and Indonesian releases demonstrate that the recipe
can be retrained with separate teachers, weights, and frontends, but they do
not yet have comparable cross-language perceptual tests.  The automatic
English gate contains only 24 sentences, and learned predictors cannot replace
a controlled MOS study.  We also lack confidence intervals and energy
measurements.  Finally, the ESP32-S3 is the only measured real-time target;
the Arm and vector RISC-V results are projections or emulation checks.  The
standalone on-chip eSpeak frontend introduces another practical gap: it raises
end-to-end WER from 15.8\% with desktop eSpeak to 18.5\%, owing to differences
in frontend version and text normalization.

\section{Conclusion}
\label{sec:concl}

The 567{,}008-parameter, 679{,}832-byte graph demonstrates that a complete
neural TTS path can run at $0.22\times$ real time on a general-purpose
ESP32-S3 without an NPU.  It also makes the cost of that deployment clear: on
unseen text it scores 2.54 SCOREQ, well below its Kristin teacher.  In the
controlled Kristin comparison, most of the remaining loss comes from decoder
capacity.  Separately trained Amy packages show the quality available at a
larger budget, reaching 4.13 with 1{,}454{,}284 parameters and 4.16 with
1{,}834{,}380.

Under the boundary used here---external G2P excluded, all inference-time
neural tensors included, and physical MCU timing required---this is, to our
knowledge, the smallest complete neural TTS graph demonstrated in real time on
a general-purpose microcontroller without a neural accelerator.  This is an
empirical deployment claim, not a claim about the smallest possible model.
The explicit latent interface and the duration, latent, waveform, adversarial,
and joint objectives provide a
repeatable path from each VITS teacher to its students.  Evaluation on diverse
text, followed by phoneme-level analysis when listening reveals a defect,
keeps aggregate scores from concealing failure.  Just as importantly, the
released artifacts preserve the association among each teacher, its frontend,
its student weights, and the hardware on which that student was measured.

\section*{Disclosure}
Generative AI tools assisted drafting and organizing experiment logs into
tables.  All experiments, code, numerical results, and claims were designed,
executed, and verified by the author, who takes full responsibility.

\IEEEtriggeratref{12}
\bibliographystyle{IEEEtran}
\bibliography{refs}

\end{document}